\documentclass[useAMS,usenatbib]{mn2e}

\usepackage{graphicx}
\usepackage[
    a4paper,
    plainpages=false,
    breaklinks=true,
    bookmarks=true,
    bookmarksopen=true,
    bookmarksopenlevel=1
]{hyperref}

\def\lesssim{\mathrel{\hbox{\rlap{\hbox{\lower4pt\hbox{$\sim$}}}\hbox{$<$}}}}
\def\gtrsim{\mathrel{\hbox{\rlap{\hbox{\lower4pt\hbox{$\sim$}}}\hbox{$>$}}}}

\title[Primordial Globular Clusters, XRBs \& Reionisation]
      {Primordial Globular Clusters, X-Ray Binaries \& Cosmological Reionisation}
      \author[C. Power, G. A. Wynn, C. Combet, M. I. Wilkinson]
	     { C. Power\thanks{chris.power@astro.le.ac.uk},
               G. A. Wynn, C. Combet \& M. I. Wilkinson\\
               Department of Physics \& Astronomy, University of 
               Leicester, Leicester LE1 7RH, United Kingdom}
\begin{document}

\date{}

\pagerange{\pageref{firstpage}--\pageref{lastpage}} \pubyear{2009}

\maketitle

\label{firstpage}

\begin{abstract}
Globular clusters are dense stellar systems that have typical ages of $\sim
13$ billion years, implying that they formed during the early epochs of
galaxy formation at redshifts of $z \gtrsim 6$. Massive stars in newly formed
or primordial globular clusters could have played an important role during the 
epoch of cosmological reionisation ($z \gtrsim 6$) as sources of energetic, 
neutral hydrogen ionising UV photons. We investigate whether or not these 
stars could have been as important in death as sources of energetic X-ray
photons as they were during their main sequence lives. Most massive stars 
are expected to form in binaries, and an appreciable fraction of these (as 
much as $\sim 30\%$) will evolve into X-ray luminous ($L_X \sim 10^{38}$
erg/s) high-mass X-ray binaries (HMXBs). These sources would have made a 
contribution to the X-ray background at $z \gtrsim 6$. 

Using Monte Carlo models of a globular cluster, we estimate the 
total X-ray luminosity of a population of HMXBs. We compare and contrast 
this with the total UV luminosity of the massive stars during their main 
sequence lives. For reasonable estimates, we find that the bolometric 
luminosity of the cluster peaks at $\sim 10^{42}$ erg/s during the first 
few million years, but declines to $\sim 10^{41}$ erg/s after 
$\sim 5$ million years as the most massive stars evolve off the main
sequence. From this time onwards, the total bolometric luminosity is
dominated by HMXBs and falls gradually to $\sim 10^{40}$ erg/s after $\sim 50$
million years. Assuming a power-law spectral energy distribution for the
HMXBs, we calculate the effective number of neutral hydrogen ionisations per
HMXB and show that HMXBs can be as important as sources of ionising radiation 
as massive stars. Finally we discuss the implications of our
results for modelling galaxy formation at high redshift and the prospects of
using globular clusters as probes of reionisation.
\end{abstract}

\begin{keywords}
globular clusters: general -- galaxies: formation -- X-rays: binaries -- cosmology:theory
\end{keywords}

\section{Introduction}
\label{sec:intro}

Hydrogen is the most abundant element in the Universe and it is fundamental to
galaxy formation, representing the principal raw material from which stars 
form. Approximately 16\% of the matter content of the Universe at present is
baryonic \citep[cf.][]{2007ApJS..170..377S}, of which about 0.34\% is cold gas
\citep[most of which is atomic and molecular hydrogen; cf. Table 1 of
][]{2004ApJ...616..643F}. From these numbers we conclude that the bulk of 
cosmic hydrogen is ionised at present, yet there must have been a period early 
in the history of the Universe when the bulk of hydrogen was neutral. This
is supported by a range of observational data that provides strong and
compelling evidence that the Universe underwent an ``Epoch of Reionisation'' 
that was complete by $z \sim 6$, approximately 1 billion years after the Big Bang 
\citep[e.g.][]{2001AJ....122.2850B,2007ApJS..170..377S}. During this period
the cosmic abundance of neutral hydrogen declined dramatically, ``re-ionised'' 
by a background of ionising UV and X-ray radiation whose build-up was very
likely linked to the formation of the first generations of stars and galaxies 
\citep[e.g.][]{2007RPPh...70..627B}. 

Understanding the precise nature of the sources of this ionising radiation
background remains an important yet largely unsolved problem facing modellers
of galaxy formation. It is important because reionisation is expected 
to have had a dramatic impact on galaxy formation. The presence of a 
photo-ionising background can inhibit the collapse of baryons onto low-mass
dark matter haloes \citep[e.g.][]{1992MNRAS.256P..43E,1996ApJ...465..608T} and 
suppress radiative cooling and subsequent star formation within dark matter 
haloes \citep[e.g.][]{2002MNRAS.333..156B}. Reionisation has been invoked to reconcile the 
apparent disparity between the observed abundance of satellite galaxies in the 
Local Group with the abundance inferred from simulations of galaxy halo 
formation within the favoured Cold Dark Matter framework
\citep[e.g.][]{2002MNRAS.333..177B}. In addition, there is good reason to 
believe that reionisation was also important in shaping the faint end 
of the luminosity function \citep[e.g.][]{2003MNRAS.344..835B}, the clustering 
of galaxies \citep[e.g.][]{2007MNRAS.382..921W}, the low-mass end of the HI
mass function, and perhaps the spatial distribution of globular clusters 
\citep[e.g.][]{2006MNRAS.368..563M}.

However, despite its importance for galaxy formation, reionisation remains a
largely unsolved problem because very little is known about the sources of
re-ionising radiation. This is because it is technically challenging 
to observe directly the Universe at $z \gtrsim 6$, and will remain so until 
the advent of next generation instruments such as LOFAR
\citep[e.g.][]{2003NewAR..47..405R,2005MNRAS.360L..64Z} and
JWST \citep[e.g.][]{2006NewAR..50..113W,2008arXiv0809.3926H}. Therefore 
much of our understanding of the properties of potential sources of 
reionising radiation comes from a combination of cosmological simulations 
\citep[e.g. ][]{2004MNRAS.352..547R,2004MNRAS.350...47S} and
(semi-)analytical galaxy formation modelling 
\citep[e.g.][]{2006MNRAS.369.1055B,2007MNRAS.375.1269Z}, whose predictions can
be tested against limits inferred from the spectral energy density of the 
radiation background at high redshifts \citep[e.g.][]{2004ApJ...613..646D} and 
the electron-scattering optical depth of the cosmic microwave background 
$\tau_e$ \citep[e.g.][]{2008ApJ...685....1S}. Not unexpectedly, 
UV luminous massive stars
\citep[e.g.][]{2003ApJ...588L..69W,2004MNRAS.350...47S,2008ApJ...684....1W}
and systems in which there is accretion onto X-ray luminous intermediate mass
and supermassive black holes
\citep[e.g.][]{2004MNRAS.352..547R,2005MNRAS.357..207R,2007MNRAS.375.1269Z}
have been suggested as likely sources of the ionising radiation background. 
However, there are many unanswered questions about the nature and origin of 
these ionising sources, and without direct observations of the galaxy 
population at $z \gtrsim 6$ any answers we have will be largely speculative.

There are fossil records of galaxy formation in the high redshift Universe
in our own cosmic backyard. Our Galaxy and others in the Local Group have
large populations of globular clusters, the bulk of which are metal poor 
\citep[cf. Figure 2 of][]{2006ARA&A..44..193B} and old ($\sim 13$ billion
years old). They appear to be relatively simple systems -- their stellar 
populations appear to be generally coeval, forming in one or more 
bursts, and many aspects of their dynamical evolution have been studied in 
detail by direct $N$-body methods \citep[see, for example,][Praagman, Hurley
  \& Power 2008]{2007ApJ...665..707H}. 

This apparent simplicity has led globular clusters to be used increasingly 
as probes of high redshift galaxy formation \citep[see, for example, the
  review of ][]{2006ARA&A..44..193B} because they may, in principle, tell us 
about the conditions in which galaxy formation proceeded at early times. 
For example, the inferred ages of old metal poor globular clusters implies 
that these systems formed at a time when the Universe was undergoing 
reionisation. Reionisation is expected to quench star formation and therefore
globular cluster formation, and so it has been suggested that present-day 
spatial distribution of metal poor globular clusters around galaxies can be 
used to measure the redshift of cosmological reionisation 
\citep[e.g.][]{2006MNRAS.368..563M,2006MNRAS.372.1019B}. 

However, globular clusters themselves may be sources of ionising 
radiation and they could be potentially important for cosmological 
reionisation. \citet{2002MNRAS.336L..33R} (hereafter R02) has pointed out 
that massive stars in primordial globular clusters at high redshifts 
are extremely luminous at UV wavelengths, which means that they could have
been effective ionising sources of neutral hydrogen provided the 
radiation could escape freely from its source 
\citep[e.g.][]{2000ApJ...542..548R,2006MNRAS.369.1055B}. Globular clusters 
tend to reside on the outskirts of galaxies at the present epoch and if this
was the case at early times then the fraction of emitted ionising UV photons
 that can escape without being scattered or absorbed would have been significant. 
R02 estimated that massive stars in a primordial globular cluster could emit UV
photons sufficiently energetic to ionise atomic hydrogen ($E_{\gamma} \geq
13.6$ eV) at a rate of $\dot{N}_{\gamma}\sim 3 \times 10^{53}$ s$^{-1}$ over
the first 4 million years; this is equivalent to a mass of hydrogen of
between $\sim 10^{9} \rm M_{\odot}$ and $\sim 10^{10} \rm M_{\odot}$,
depending on the local recombination rate \citep[cf.][]{2004ApJ...613..646D}.

Massive stars in the local Universe tend to form with one or more companions
\citep[e.g.][]{1996A&A...315..384R,1998AJ....115..821M,2004MNRAS.351..617D},
and this was likely to be the case in primordial globular clusters. Massive 
binaries may evolve into high mass X-ray binaries 
(hereafter HMXBs) once the more massive star collapses to form a compact object,
either a neutron star or a black hole. This compact object accretes material
from its companion, either via a stellar wind or Roche lobe overflow 
(RLOF), which results in X-ray emission. The HMXB phase is believed 
to occur soon after the first compact object is formed and its duration is
limited by the main sequence lifetime of the secondary, typically $\sim 10^7$
years. HMXBs in our Galaxy are observed to emit strongly in X-rays,
with typical luminosities of $L_X\sim 10^{35}-10^{38}$~erg~s$^{-1}$ 
\citep[cf.][]{2006A&A...455.1165L}, and there
is good reason to believe that HMXBs in primordial globular clusters at high
redshifts would have been as luminous, if not more so \citep[e.g.][]{2006Ap&SS.304..279D}.

This is very interesting because it suggests that massive stars in globular 
clusters could have been important as sources of both ionising UV radiation 
and X-rays during the epoch of cosmological radiation. Certainly, primordial
globular clusters would have formed an abundance of UV luminous massive stars
that would have had a profound impact on cold star-forming gas in their
immediate surroundings. However, could primordial globular clusters have
formed sufficient numbers of HMXBs to make an interesting contribution to the
ionising X-ray background during the epoch of reionisation? Whereas UV 
photons are readily
absorbed by neutral hydrogen, X-rays are much more penetrating (the
photo-ionising absorption cross section of neutral hydrogen decreases with
photon energy roughly as $E_{\gamma}^{-3}$) and can escape into the 
inter-galactic medium. 

In this paper we consider what fraction of massive stars in primordial
globular clusters must
evolve into HMXBs for these sources to make a significant contribution to the 
X-ray ionising background. We address this question using a Monte Carlo model 
of a primordial globular cluster of mass $10^6 \rm M_{\odot}$. We assume that 
all massive stars form in binaries \citep[see, for example,][]
{2006MNRAS.370.2079D} and explore what fraction of these binaries must 
eventually form HMXBs to be an important source of reionising photons. 

The layout of the paper is as follows; in \S~\ref{sec:hmxbs}, we
discuss the factors that determine the fraction of massive binaries that
survive to evolve into HMXBs ($f_{\rm sur}$) and, provided they survive, their
X-ray luminosities. In \S~\ref{sec:model} and \S~\ref{sec:reion}, we
describe our cluster model in some detail and we compute the effective rate 
at which neutral hydrogen could be ionised by UV photons from main sequence 
stars and by X-ray photons from HMXBs. Finally, in
section~\ref{sec:discussion} we discuss the implications of our results for 
modelling of high redshift galaxy formation and the use of globular clusters 
as probes of cosmological reionisation.

\section{The Ionising Power of Young Globular Clusters}

In the following subsections, we consider the factors that regulate the
formation of high mass X-ray binaries (\S~\ref{sec:hmxbs});
we present the details of our globular cluster model and estimate the total
X-ray luminosity as a function of time (\S~\ref{sec:model}); and we determine
what fraction of this luminosity is available to ionise neutral hydrogen 
(\S~\ref{sec:reion}).

\subsection{The survival fraction $f_{\rm sur}$}
\label{sec:hmxbs}

Not all massive binaries will evolve to become HMXBs. If a massive binary is 
to survive and become a HMXB, then the stars in the binary must not merge 
during the more massive stars' main sequence (MS) and post-MS evolution, and
the binary must survive the supernova of the more massive star. 
Whether or not a massive binary merges during the MS 
lifetime of the more massive star is determined principally by the details of 
stellar evolution at low metallicities and the initial binary separation. If
the binary survives without merging, then the
issue of whether or not the binary survives the first supernova depends on a 
number of factors, including the precise mass of the more massive star, the 
fraction of mass lost during the supernova and the kick 
velocity imparted during the supernova. However, binary evolution 
in globular clusters at very low metallicities is not particularly well 
understood. 
Examples of complicating factors include the fact that lower metallicity stars 
are expected to retain more of their mass because stellar winds are
inefficient at low metallicities; therefore these stars will be more massive 
at the end of their main sequence lives and will form black holes rather than 
neutron stars \citep[cf.][]{2003fthp.conf....3H}. This is consistent with the 
expectation that the ratio of black hole to neutron star remnants is much 
higher at low metallicities \citep[e.g.][]{2008arXiv0803.3161H}. Also, less
mass is lost during the supernova when the remnant is a black hole rather than 
a neutron star, and black holes may suffer less violent natal kicks.
These arguments suggest that binaries that survive to become HMXBs in
primordial globular clusters are more likely to contain black holes, 
which suggests in turn that they could be more X-ray luminous than 
typical HMXBs in the local Universe, which are dominated by neutron star 
systems. 

To avoid some of the uncertainty surrounding the stellar physics in primordial
globular clusters we introduce a survival fraction $f_{\rm sur}$ which
indicates the likelihood of an individual binary becoming an HMXB. For each 
primordial binary, we determine whether or not the binary will remain bound 
when the more massive star goes supernova by estimating the fraction of mass 
lost in the supernova explosion; if the binary remains bound, then it has a 
probability of $f_{\rm sur}$ that it will evolve into a HMXB. We find that 
typically $70\%$ of high-mass binaries become unbound following the supernova 
of the more massive star, and so at most $\sim 30\%$ of the initial binary 
population will evolve to become HMXBs. We note that those that survive tend 
to host black holes rather than neutron stars. It is the fraction 
$f_{\rm sur}$ of this remaining $\sim 30\%$ of bound binaries which go on to 
become HMXBs. This approach allows us to investigate the potential 
effectiveness of HXMBs as sources of reionising radiation in a robust manner, 
even if the factors that determine the value of $f_{\rm sur}$ are not well 
understood, we can speculate on what the most important determinants of 
$f_{\rm sur}$ are likely to be.

\subsection{Modelling HMXBs in Young Globular Clusters}
\label{sec:model} 

We wish to assess the potential importance of the HMXB population in
globular clusters for the X-ray background at high redshifts, and so we
require a model of the stellar population in a typical globular cluster.
We use a Monte Carlo model of a globular cluster of $10^6$ stars and follow
the evolution of its population of massive stars over its first 100 million 
years, through their main sequence lives and into the HMXB phase. The 
main features of our model can be summarised as follows;

\begin{itemize}

\item {\bf The Initial Mass Function.} We adopt three different IMFs -- those
  of \cite{1955ApJ...121..161S}, \cite{2001MNRAS.322..231K} and 
  \cite{2001ApJ...554.1274C} -- with lower and 
  upper mass cut-offs of $0.01 {\rm M_{\odot}}$ and $100 {\rm M_{\odot}}$ 
  respectively. The most important prerequisite for a HMXB to form is that the 
  primary is either a neutron star or a black hole, which sets a lower mass 
  limit of $M \gtrsim 8 \rm M_{\odot}$ for the primary \citep[cf. Figure 1 
  of][]{2003ApJ...591..288H}; this is the threshold for neutron star 
  formation. The mean number of stars with masses in excess of 8 
  $\rm M_{\odot}$ and 20 $\rm M_{\odot}$ (the threshold for black hole 
  formation) are for Salpeter 2621 (753), for Kroupa 11126 (3174) and for 
  the top-heavy Chabrier 131935 (27024).
  
\item {\bf Binary Formation.} We assume that all massive stars form in
  binaries. Initial binary orbital
  parameters are assigned following the approach of \cite{2006MNRAS.370.2079D} 
  -- companion masses are drawn from a uniform distribution between 0.1 and
  100 $\rm M_{\odot}$ and orbital periods are distributed uniformly in
  logarithm between 1 and $10^4$ days.
   
\item {\bf Massive Star Lifetimes.} Massive stars have short main sequence 
  lifetimes, the duration of which are determined principally by their
  metallicity. We obtain estimates for
  these lifetimes using the results of \cite{2008arXiv0807.0573E}, 
  \cite{1993A&AS..102..339S} and \cite{2000A&A...361..101M} for Z=0, 0.008 and 
  0.02 (i.e. solar metalicity) respectively. 

\item {\bf HMXB Formation.} When the more massive star in the binary reaches 
  the end of its main
  sequence lifetime, we assume that it goes supernova and forms either a
  neutron star or black hole. The remnant's mass -- which is determined by the
  mass of the star at the end of its main sequence life -- is estimated
  following Figure 3 of \citep[][]{2003ApJ...591..288H}. We calculate revised
  binary parameters (i.e. semi-major axis and period) although there are
  additional factors which complicate matters (e.g. velocity kick imparted by 
  the supernova). If the system loses more than half its mass in the
  supernova, it becomes unbound and we remove it from the list of potential HMXB
  candidates; this removes $\sim 70\%$ of binaries from consideration
  if we assume a Kroupa IMF. We then draw a fraction $f_{\rm sur}$ of the 
  remaining $\sim 30\%$ at random and consider them as HMXBs.

\item {\bf HMXB Luminosities.} If a binary survives and forms an HMXB, we
  estimate its X-ray luminosity. Although binary parameters are likely to play 
  a role in determining the X-ray luminosity \citep[see, for example, 
  ][]{2006MNRAS.370.2079D}, we take a simpler approach, drawing luminosities
  from a Weibull distribution with a peak fixed at $L_X \sim 10^{38}$ erg/s 
  but preventing a HMXB from accreting at greater than its
  Eddington limit. This sets an upper limit of approximately 
  $L_X \simeq 1.26 \times 10^{38} (M/\rm M_{\odot})$ erg/s. We
  note that this is consistent with the luminosities of compact X-ray
  sources in nearby galaxies whose X-ray binary populations are dominated by
  HMXBs \citep[cf. Figure 1 of ][]{2004NuPhS.132..369G}; see
  \S~\ref{sec:discussion} for further discussion.
  
\item {\bf HMXB Lifetimes.} We assume that HMXBs are active until the
  companion star evolves off the main sequence and goes supernova. 

\end{itemize}

\noindent Note that although we investigate the sensitivity of our model's
results to the choice of IMF (i.e. Salpeter, Kroupa or Chabrier), for clarity
we concentrate on models that assume the Kroupa IMF only.

\begin{figure}
  \includegraphics[width=8cm]{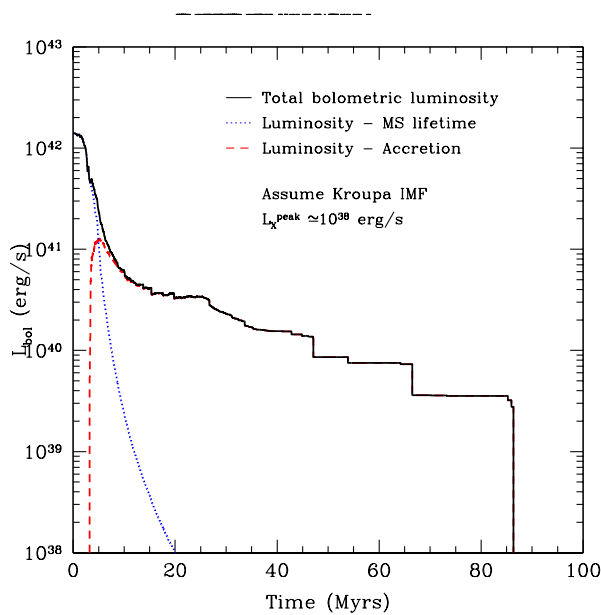}
  \caption{{\bf Variation of Total Bolometric Luminosity with Time}. The light
  dotted curve corresponds to the contribution of massive stars during their
  main sequence lives, while the dashed curve corresponds to the contribution
  of the HMXBs. The solid curve indicates the net luminosity.}
  \label{fig:lumx_time}
\end{figure}

In Figure~\ref{fig:lumx_time} we show how the total bolometric luminosity of
our model cluster varies with time, where we have assumed that $f_{\rm
  sur}$=1, i.e.  all of the $\sim 30\%$ of the initial massive binary population
that remain bound after the primary goes supernova. During
the first few million years the luminosity is dominated by the contribution
from massive stars, but this declines rapidly as these massive stars evolve
off the main sequence and in some cases become HMXBs. The HMXB contribution 
grows rapidly once the most massive main sequence stars go supernova (after 
$\sim 3.4$ million years for our assumed upper mass cut-off) and dominates 
the total luminosity from $\sim 5$ million years onwards. 

The total bolometric luminosity for the cluster peaks at $\sim 10^{42}$ erg/s
during the first few million years, but it declines rapidly and has dropped
to $\sim 10^{41}$ erg/s after $\sim 5$ million years. The luminosity during
this period is dominated by the contribution of the massive stars during their
MS lives. After $\sim 5$ million years the
decline becomes more gradual, falling by a factor of $\sim 30$ over the next
$\sim 80$ million years, during which time the luminosity is dominated by 
the contribution of HMXBs. From this we may conclude that massive stars are
indeed energetic sources of radiation during their MS lives and
in terms of their total bolometric luminosity they produce as much energy as
HMXBs, whose contribution extends over a much longer period.

\subsection{Implications for Cosmological Reionisation}
\label{sec:reion}

\begin{figure}
  \includegraphics[width=8cm]{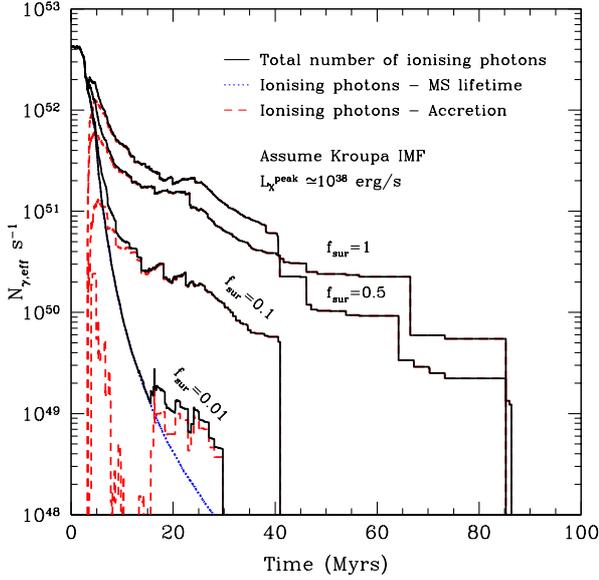}
  \caption{{\bf Variation of Rate of Emission of Hydrogen Ionising Photons with Time}.}
  \label{fig:nphot_ion}
\end{figure}

The total bolometric luminosity is an interesting number, but what fraction of
this energy is available to ionise neutral hydrogen? We compute the total number
of ionising photons emitted per second by both massive stars during their main
sequence lives and by HMXBs and show its variation with time over the first
$100$ million years of the globular cluster's life in Figure~\ref{fig:nphot_ion}.
We model the ionising luminosity of massive stars using the {\small
  Starburst99} code \citep[cf.][]{1999ApJS..123....3L} for the specific case
of a cluster of $10^6$ stars, assuming that all of the stars were formed in an
instantaneous burst and Geneva tracks with high mass loss for metallicities of
$Z$=0.001, 0.008 and 0.02 (i.e. solar). The ionising luminosity of HMXBs is
calculated by assuming a spectral energy distribution that follows a simple power-law
$F(E) \propto E^{-\alpha}$ between lower and upper energy cut-offs of 
$E_{\rm min}$=0.1 eV and $E_{\rm max}$=10$^6$ eV respectively. The total
energy liberated during accretion is then 
\[ 
E_{\rm tot} = (A/(2-\alpha))(E_{\rm max}^{2-\alpha}-E_{\rm  min}^{2-\alpha})
\] 
\noindent and so given the luminosity it is straightforward to compute the normalisation
constant $A$ for a given HMXB. We estimate the effective number of hydrogen ionising
photons to be

\begin{equation}
  \label{eq:fofe}
        { N_{\gamma,\rm eff} = (13.6\rm eV)^{-1}\int^{E_{\rm lim}}_{13.6 eV} F(E)dE}
\end{equation}

\noindent where $E_{\rm lim}$ is the energy above which the mean free path of
photons becomes of order the Hubble length (i.e. atomic hydrogen becomes
transparent to hard X-rays). We obtain $E_{\rm lim}$ by requiring that 
$\sigma(E_{\rm lim})=(H(z)/c)/n(z)$ where $\sigma(E)$ is the 
ionisation cross section of neutral hydrogen, $H(z)$ is the Hubble parameter at 
redshift $z$, $n(z)$ is the mean baryon density and $c$ is the speed of
light. For the redshifts in question ($z \gtrsim 6$) we find that $E_{\rm lim}
\gtrsim$ 1keV, although the precise value of $E_{\rm lim}$ has relatively
little impact on $N_{\gamma}$. Equation~\ref{eq:fofe} makes the simple
assumption that an energetic photon can ionise multiple hydrogen atoms, and so
we treat secondary electrons that ionise hydrogen atoms as effective photons.

\begin{figure}
  \includegraphics[width=8cm]{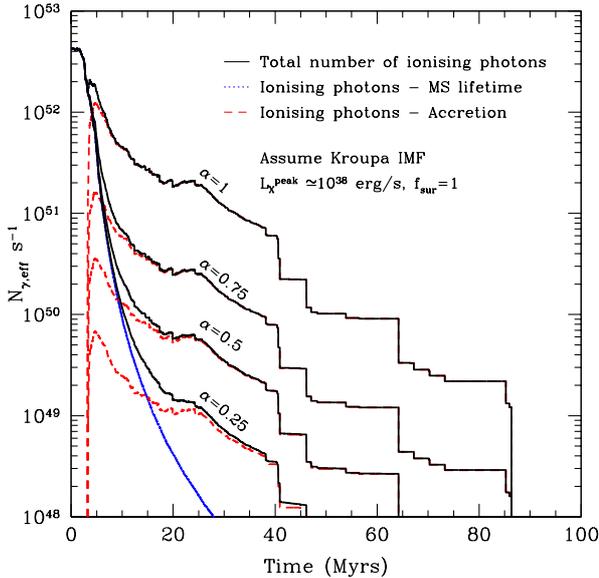}
  \caption{{\bf Ionsing Power and Dependence on Assumed Hardness of Energy Spectrum}.}
  \label{fig:alpha_dependence}
\end{figure}

In Figure~\ref{fig:nphot_ion} we show how the rate of emission of ionising
photons from massive stars (dotted curve), from HMXBs (dashed curve) and from
ionising sources regardless of their nature (solid curve) varies with time,
assuming $F(E) \propto E^{-1}$, which allows us to gauge the
effectiveness of HMXBs as ionising sources as a function of $f_{\rm sur}$.
At their peak, massive stars produce hydrogen ionising photons at
a rate of $\sim 4 \times 10^{52} s^{-1}$ during the first few million years of
the globular clusters life, but this declines rapidly. We note that the peak
value is a factor of $\sim 10$ lower than is quoted by
\citet{2002MNRAS.336L..33R} but this reflects different assumptions he
made about, for example, metallicity ($Z \sim Z_{\odot}$) and the IMF
(Salpter). As one expects, the
effective number of ionising photons produced by HMXBs depends strongly on
what one assumes for $f_{\rm sur}$ and as $f_{\rm sur}$ decreases, so too does
the strength and duration of the HMXB population as a source of (effective) 
ionising photons. For survival fractions $f_{\rm sur}$ between $50\%$ and
$100\%$ the peak rate at which (effective) ionising photons are emitted is between 
$\sim 5 \times 10^{51} s^{-1}$ and $10^{52} s^{-1}$, but this declines
gradually and after 30 million years the rate is $\sim 10^{51} s^{-1}$.

It is important to note that we have assumed a particular hardness for our
energy spectrum ($\alpha=1$); in Figure \ref{fig:alpha_dependence} we show how
our results depend on $\alpha$, with lower values of $\alpha$ corresponding to
harder energy spectra. As energy spectra become harder (i.e. as $\alpha$ decreases)
HMXBs become less effective as ionising sources; this is
unsurprising because the proportion of energetic photons increases as the
hardness of the source increases, with a corresponding decrease in the number
that can be absorbed by neutral hydrogen. This provides an interesting
additional constraint on the precise nature of HMXBs in primordial globular
clusters, if they are to be considered as important contributors to 
cosmological reionsiation.

Figure~\ref{fig:nphot_ion} suggests that the hydrogen ionising power of a
single young globular cluster is at its most effective when its massive star 
population is still on the main sequence. However, the ionisation cross
section of neutral hydrogen is a strong function of photon energy, decreasing
roughly as $E^{-3}$, so that the mean free path of a UV photon is much shorter
than that of an X-ray photon. Therefore we expect UV photons to be most
effective ionising the relatively high density surroundings of galaxies while
X-ray photons can escape unscathed from the galaxy to potentially ionise a 
much larger volume \citep[e.g.][]{2008MNRAS.387..158R}. Because UV photons are
absorbed in higher density surroundings, we expect recombination to be
important and so several UV photons may be required to ionise a single atom of
hydrogen. This has been noted by \citet{2004ApJ...613..646D}, who estimate
that $\sim 10$ UV photons are required to ionise a single hydrogen atom,
compared to $\lesssim 1$ X-ray photon. This is very interesting because it 
implies that the HMXB population is likely to be as important as massive 
stars as a source of hydrogen ionising radiation. Inspection of 
Figure~\ref{fig:nphot_ion} shows that even if only a few UV photons (rather
than the $\sim 10$ estimated by \citet{2004ApJ...613..646D}) are required to
ionise a single hydrogen atom, the relative importance of HMXBs is boosted
dramatically and these systems now dominate as sources of ionising
radiation. Provided the survival fraction $f_{\rm sur} \gtrsim 0.5$ and
$\alpha \simeq 1$, we would
expect a single globular cluster of $10^6$ stars to ionise of order $\sim 5
\times 10^7 \rm M_{\odot}$ of neutral hydrogen during its first $\sim 100$
million years. How precisely one interprets this number is not straightforward
because UV photons are absorbed locally whereas X-ray photons contribute
to a global X-ray background. Nevertheless, it suggests that the HMXB
population that we expect to be present in newly formed globular clusters at
high redshifts could have been as important as massive stars for cosmological
reionisation \citep[cf.][]{2002MNRAS.336L..33R}.

\section{Conclusions}
\label{sec:discussion}

Globular clusters are old, dense and relatively simple stellar systems whose
dynamical properties and evolution are well understood. These properties have 
led to increasing interest in globular clusters as probes of the conditions 
under which galaxies formed at high redshifts 
\citep[cf.][]{2006ARA&A..44..193B}. The study of 
\citet{2002MNRAS.336L..33R} has suggested that massive stars in
primordial globular clusters could have made an important contribution
to cosmological reionisation as extremely luminous sources of UV photons.
We have explored the contribution that these massive stars could have made to
cosmological reionisation once they evolved off the main sequence, by assuming 
that a fraction formed in binaries that evolved into HMXBs. 

Using a Monte Carlo model of a primordial globular cluster, we have
investigated the conditions under which a sufficient number of HMXBs form 
to have a meaningful impact on the ionising power of the cluster.
We assume that all massive stars formed in binaries and we consider 
only those binaries that remain bound once the more massive star goes
supernova, typically $\sim 30\%$ of massive binaries. Of this $\sim 30\%$, we
assume that a fraction $f_{\rm sur}$ survives to form HMXBs. Note that this
is a conservative estimate because we have neglected the effect of mass loss 
over the MS lifetime of the more massive star.

Our results show that we require $f_{\rm sur} \gtrsim 0.5$ 
if HMXBs are to be as effective as MS massive stars as 
sources of ionising radiation. Assuming
$f_{\rm sur} \sim 1$ and a typical X-ray luminosity of $L_X \sim 10^{38} \rm
erg/s$, we find that HMXBs produce ionising photons at an effective rate of 
$\gtrsim 10^{51} s^{-1}$ over the first 40 million years of the cluster's
life. This is comparable to the rate at which massive stars produce
ionising UV photons during the first few million years. Note however that 
result is sensitive to the assumed hardness of the energy spectrum of the 
HMXBs -- the harder the energy spectrum (i.e. the smaller the power-law
exponent $\alpha$), the less effective HMXBs are as ionising sources. 

By its nature, our modelling is speculative -- we know relatively little about
the conditions in primordial globular clusters, and little about binary
evolution at low metallicities. Yet, we know that many globular
clusters formed at high redshifts -- at least as many as we can observe in the
present day Universe -- and we have a good understanding of the initial mass
function and metallicities of the stars. This allows us to estimate how many
massive stars there might have been and what their typical lifetimes were.
Our analysis indicates that under the right conditions, HMXBs in primordial globular
clusters can be effective sources of ionising radiation. If these
conditions are satisfied (high survival rate $f_{\rm sur}$, softer energy
spectra) then it reveals that accretion power onto neutron stars and
stellar mass black holes can make a contribution to reionisation.

We have assumed that the typical X-ray luminosity of an HMXB in 
a primordial globular cluster is $L_X \sim 10^{38} \rm erg/s$. This is more 
luminous than is typical of local HMXBs in our Galaxy and the neighbouring 
Large and Small Magellanic Clouds \citep[e.g. there is a single source with 
$L_X \sim 10^{38} \rm erg/s$ in the SMC; cf. ][]{2005MNRAS.362..879S},
but these environments are atypical of the kind we might expect to find
primordial globular clusters forming in. Instead, if we consider HMXB 
populations in galaxies that have undergone recent merging activity and/or 
show high star formation rates, we find that $L_X \sim 10^{38} \rm erg/s$ is 
typical \citep[cf. Figure 1 of][]{2004NuPhS.132..369G}. For example, the 
HMXB populations in recent galaxy mergers such as ``The Antennae'' 
\citep[cf.][]{2001ApJ...554.1035F} and ``The Cartwheel''
\citep[cf.][]{2004A&A...426..787W} have X-ray luminosities
in excess of $L_X \sim 10^{38} \rm erg/s$. This is interesting because, as we
discuss below, young star clusters in such galaxy mergers are likely to be
local analogues of primordial globular clusters, and suggests that our assumed
X-ray luminosities are reasonable.

It is less straightforward to assess whether or not HMXBs form in the
numbers we might expect based on our model in young star clusters in ``The
Antennae'' or ``The Cartwheel''. The luminosity functions presented 
in \citet{2004NuPhS.132..369G} and \citet{2004A&A...426..787W} can provide us 
with some insight, however.

We find that approximately 70\% of massive binaries are disrupted before
they can become HMXBs in our model. Therefore we expect at most 786 HMXBs for 
a Salpeter IMF and 952 HMXBs for a Kroupa IMF to survive in a 1e6 solar 
mass cluster, and we introduce a factor $f_{sur} \leq 1$ to parameterise the
uncertainty as to what fraction of this surviving population evolve into
HMXBs. In the most optimistic case, $f_{sur}=1$. In this case, for our adopted 
probability distribution, approximately 10\% will have luminosities in excess 
of $L_X \sim 10^{38}$ erg/s -- corresponding to between 80 and 100 HMXBs in 
a single cluster. These numbers will scale in proportion to $f_{sur}$, 
so $f_{sur}=0.1$ implies between 8 and 10 HMXBs more luminous than 
$L_X \sim 10^{38}$ erg/s

Of order 50 HMXB candidates are known in ``The Antennae''. Without knowing
the detailed spatial distribution of these candidates, it is difficult to do a
straightforward comparison -- the sources may be associated with a single
cluster, in which case $f_{sur}=1$ would seem to be favoured, or they could be
associated with 10 clusters in which case $f_{sur}=0.1$ would be
favoured. Clearly, this is a very uncertain game to play and would require
more detailed modelling (e.g. taking observed distribution of SSCs in the
Antennae, populating them with HMXBs and making mock observations), but the
numbers do not seem unreasonable. 

The crucial factor here is $f_{sur}$. For HMXBs in primordial globular
clusters to be an important source of ionising radiation, we require $f_{sur}$
to be of order unity. We expect $f_{sur}$ to depend on many factors, one of 
which will be metallicity; $f_{sur}$ may decrease as stars become more metal 
rich, and so it may very well be the case that $f_{sur}=0.1$ rather than $\sim$
1 in a system such as the Antennae. Establishing how different factors impact 
upon $f_{sur}$ and how these factors may depend on redshift is something that 
we would like to investigate in future work, but at present the discussion 
is necessarily speculative. \\

We have focussed on the ionising power of HMXBs in primordial globular
clusters, but it is important to note that the longer mean free path of
X-rays compared to UV means that HMXBs in galactic discs could have been as 
effective as HMXBs in globular clusters. We focus on globular clusters 
because we have a reasonable understanding of what the IMF could have 
been, and therefore we have a reasonable idea of how many HMXBs were 
likely to form. In contrast, we know very little about the initial mass 
function in star-forming galactic discs at high redshifts, but this is not 
to say that HMXBs in discs should be discounted. Rather it suggests that 
stellar mass black holes could make an interesting contribution to 
reionisation and merit further study. The contribution of globular cluster 
HMXBs to the reionising background considered in this paper could be 
considered as a lower limit.  

More broadly our analysis strongly indicates that globular clusters could have 
been important ionising sources in the high redshift Universe. This has
interesting implications for galaxy formation at high redshifts and the 
use of globular clusters as probes of reionisation. As yet, we do not have a
satisfactory theory for the formation of globular clusters within a
cosmological context. Nevertheless, if they were to be effective sources of 
UV radiation 
then they must have moved quickly from their formation site, presumably a 
cocoon rich in cold dense gas, to the rarified environs of the hot gaseous
halo; if this did not happen on a timescale shorter than a fraction of a 
few million years, a good fraction of their UV radiation would have been
absorbed by the cold dense gas. This would favour globular cluster forming
in gas-rich mergers akin to the ``super star clusters'' that we observe in
present day mergers \citep[e.g.][]{1995AJ....109..960W,1995Natur.375..742M}
rather than in the gas-rich discs \citep[e.g][]{2005ApJ...623..650K}. 

We have argued that ionising radiation from
the globular clusters could have been effective in ionising neutral hydrogen
surrounding their host galaxy, but this radiation could also damage cold dense
gas within the galaxy (both neutral and molecular), thus affecting the star
formation rate and potentially suppressing further star formation. 
This has particular implications for the use of globular clusters as probes of
cosmological reionisation. The argument is that the spatial distribution of
globular clusters around galaxies and in galaxy clusters can be used to
measure the epoch of reionisation -- the more centrally concentrated the
distribution of globular clusters, the earlier the epoch of reionisation.
Yet if the star formation efficiency in a galaxy is high, and consequently
the rate at which globular clusters form is high, then we might expect
feedback from early-forming globular clusters to be extremely damaging for
later-forming globular clusters \citep[e.g.][]{2006MNRAS.368..563M}. This 
would suppress the star formation and globular cluster formation efficiency. 
How could one separate this local effect from the effect of a
globally driven period of reionisation? This is a question we shall be
pursuing in future work.

\section*{Acknowledgements}

We thank the anonymous referee for their very helpful comments.
CP, CC and GW acknowledge the support of the theoretical astrophysics rolling 
grant at the University of Leicester. MIW acknowledges support from a Royal
Society University Research Fellowship.

\vspace{0.1cm} \bsp

\bibliographystyle{mn2e}

\label{lastpage}

\end{document}